# A Two-Fluid Method for Ambipolar Diffusion

David A. Tilley[1] (dtilley@nd.edu) and Dinshaw S. Balsara[1] (dbalsara@nd.edu)

**Key Words:** Diffusion, MHD, Waves, Methods: Numerical, Stars: Formation, ISM: Clouds

**Mailing Address:**
Department of Physics
225 Nieuwland Science Hall
University of Notre Dame.
Notre Dame, Indiana 46556
USA

**Phone:** (574) 631-2712
**Fax:** (574) 631-5952

# A Two-Fluid Method for Ambipolar Diffusion

David A. Tilley[1] (dtilley@nd.edu) and Dinshaw S. Balsara[1] (dbalsara@nd.edu)

[1]Department of Physics, University of Notre Dame.

**Abstract:**

We present a semi-implicit method for isothermal two-fluid ion-neutral ambipolar drift that is second-order accurate in space and time. The method has been implemented in the RIEMANN code for astrophysical fluid dynamics. We present four test problems that show the method works and correctly tracks the propagation of MHD waves and the structure of two-fluid C-shocks. The accurate propagation of MHD waves in the two-fluid approximation is shown to be a stringent test of the algorithm. We demonstrate that highly accurate methods are required in order to properly capture the MHD wave behaviour in the presence of ion-neutral friction.

## 1. Introduction

Several astrophysical systems involve gas that is partially ionized, often at very low levels of ionization (fractional ionizations $\xi=n_i/n_n \sim 10^{-6} - 10^{-8}$). The treatment of the momentum exchange between neutral and ionized fluids – ambipolar diffusion – can thus become a very demanding and interesting problem. The numerical solution of the two-fluid equation can be challenging because the force terms can be very stiff in their interaction with the ionized fluid. In order to extend numerical simulations of astrophysical systems to these physical regimes of interest, it is thus desirable to have numerical methods for two-fluid ambipolar diffusion that are stable, accurate and efficient at low ionization fractions. In this paper we demonstrate that highly accurate interpolation as well as a second-order accurate semi-implicit treatment of the source terms are both essential for accurately representing the propagation of waves in a two-fluid approximation. The two-fluid methods presented here have been implemented in the RIEMANN code for astrophysical fluid and magnetofluid problems that has previously been described in Balsara (2001, 2004).

The role of ambipolar diffusion in star formation has been studied extensively in magnetically subcritical systems (Mestel & Spitzer 1956; Mouschovias 1976a,b; Shu et al. 1987; Indebetouw & Zweibel 2000). Recent work has extended this study to the trans-critical regime (Basu & Ciolek 2004, Ciolek & Basu 2006; Kudoh et al. 2007). The rate of diffusion of magnetic field due to ambipolar diffusion is enhanced in the presence of turbulence (Fatuzzo & Adams 2002, Zweibel 2002, Li & Nakamura 2004, Nakamura & Li 2005), thus allowing the possibility for core formation on timescales much smaller than the ambipolar diffusion timescale estimated from the overall properties of the cloud. Oishi & Mac Low (2006) have suggested using a two-fluid method with a heavy-ion approximation that ambipolar diffusion does not impose a minimum mass scale for cores. In a single-fluid approximation Kudoh et al. (2007) have examined in three dimensions the interplay between ambipolar diffusion and gravity in a turbulent molecular cloud, while Nakamura & Li (2005) have done the same in a two-dimensional setup that assumes a sheetlike geometry.

The presence of ion-neutral friction in the MHD equations affects wave propagation. Langer (1978) showed that the growth rate of gravitational instabilities is strongly sensitive to the magnetic field. Pudritz (1990) showed that a spectrum of MHD waves could support a cloud on large scales, while permitting collapse on small scales. Balsara (1996) showed that slow magnetosonic waves could propagate at scales significantly smaller than the friction scale with little damping, and that at large scales gravitational collapse occurs only in slow-mode wave families. The ability to accurately track the propagation of waves at a high order of accuracy with very low ionization fractions will be very important in furthering this work.

A multi-fluid treatment of the ionized and neutral gases is desirable in the presence of shocks, as a single-fluid method cannot track the structure of a C-shock. This may be important in the presence of supersonic motion, such as that inferred from the line widths measured in molecular clouds (Larson 1981; Blitz 1993; Falgarone & Phillips 1996; Jijina, Myers & Adams 1999; Motte et al. 2003). Multi-fluid methods have been used in one-dimensional studies of C-shocks (Ciolek & Roberge 2002, Roberge & Ciolek 2007. Falle (2003) devised an implicit strategy designed to be used when the charged particle inertias can be neglected. His method was used to study MHD wave evolution in the presence of ambipolar diffusion in Lim, Falle & Hartquist (2005). Smith & Mac Low (1997) use a time-explicit method that works in multiple dimensions. Mac Low & Smith (1997) devised an implicit two-fluid method that was put to use in a multidimensional application in Oishi & Mac Low (2006). Li, McKee & Klein (2006) further examined the performance of the Mac Low & Smith (1997) method using the heavy-ion approximation to alleviate the time step restrictions imposed by the Alfvén wave speed. O'Sullivan & Downes (2006, 2007) have developed a strategy that incorporates a super-timestepping algorithm to accelerate the explicit time-steps needed to stably track the flow.

We present here an implicit-explicit predictor-corrector method that is second-order accurate in space and time for solving the two-fluid ambipolar force terms. Previous formulations of semi-implicit algorithms for two-fluid ambipolar diffusion have always been fully implicit in their treatment of the force terms, and thus first-order accurate. In the present work, we remove these assumptions. We describe in detail our solution strategy in Section 2. In Section 3 we consider four test problems that demonstrate that the method works well. Section 4 presents our conclusions.

## 2. Numerical Methods

We start from the basic equations of two-fluid isothermal ideal magnetohydrodynamics, with the addition of a friction term that couples the ionized and neutral fluids:

$$\frac{\partial \rho_i}{\partial t} + \nabla \cdot (\rho_i \mathbf{v}_i) = 0 \tag{1}$$

$$\frac{\partial \mathbf{v}_i}{\partial t} + (\mathbf{v}_i \cdot \nabla)\mathbf{v}_i = c_s^2 \nabla \ln \rho_i - \frac{1}{4\pi}\mathbf{B} \times (\nabla \times \mathbf{B}) - \nabla \Phi - \gamma \rho_n (\mathbf{v}_i - \mathbf{v}_n) \tag{2}$$

$$\frac{\partial \mathbf{B}}{\partial t} = \nabla \times (\mathbf{v}_i \times \mathbf{B}) \tag{3}$$

$$\frac{\partial \rho_n}{\partial t} + \nabla \cdot (\rho_n \mathbf{v}_n) = 0 \tag{4}$$

$$\frac{\partial \mathbf{v}_n}{\partial t} + (\mathbf{v}_n \cdot \nabla)\mathbf{v}_n = c_s^2 \nabla \ln \rho_n - \nabla \Phi - \gamma \rho_i (\mathbf{v}_n - \mathbf{v}_i) \tag{5}$$

$$\nabla^2 \Phi = 4\pi G (\rho_i + \rho_n) \tag{6}$$

where $\rho_i$ and $\rho_n$ represents the density of the ionized and neutral fluids, $\mathbf{v}_i$ and $\mathbf{v}_n$ the velocities of the ionized and neutral fluids, $\mathbf{B}$ the magnetic field, and $\Phi$ the gravitational potential. The equations have been derived in Draine (1986). Eqs. (1) to (6) can be written in a flux-conservative form which is the usual form in which they are solved on a computer. Most solution strategies for these equations that are temporally second order accurate rely on a predictor-corrector formulation. The Runge-Kutta time stepping strategy is a way of making this division into predictor and corrector steps explicit. The strategies for solving the hydrodynamics, magnetohydrodynamics and self-gravity are described in previous work (Balsara 1998a,b; Balsara & Spicer 1999a,b; Balsara 2004). We thus rewrite Eqs. (2) and (5) in a momentum-conserving form:

$$\frac{\partial \mathbf{U}_n}{\partial t} = -\partial_x \mathbf{F}_n(\mathbf{U}_n) - \partial_y \mathbf{G}_n(\mathbf{U}_n) - \partial_z \mathbf{H}_n(\mathbf{U}_n) - \mathbf{S}(\mathbf{U}_n, \mathbf{U}_i) \tag{7}$$

$$\frac{\partial \mathbf{U}_i}{\partial t} = -\partial_x \mathbf{F}_i(\mathbf{U}_i) - \partial_y \mathbf{G}_i(\mathbf{U}_i) - \partial_z \mathbf{H}_i(\mathbf{U}_i) + \mathbf{S}(\mathbf{U}_n, \mathbf{U}_i) \tag{8}$$

Where $\mathbf{U}_n$ and $\mathbf{U}_i$ are the fluid vectors for the neutrals and ions, respectively; $\mathbf{F}_n$, $\mathbf{G}_n$, $\mathbf{H}_n$ are the fluxes of the neutrals in the x, y, and z directions; $\mathbf{F}_i$, $\mathbf{G}_i$, $\mathbf{H}_i$ are the fluxes of the ions in the x, y, and z directions; and $\mathbf{S}(\mathbf{U}_n, \mathbf{U}_i) \equiv \gamma \rho_n \rho_i (\mathbf{v}_n - \mathbf{v}_i)$ denotes the source terms. For subsequent notational simplicity we denote the sum of the gradients of the fluxes acting on the two fluids by $\mathbf{L}_n(\mathbf{U}_n) \equiv -\partial_x \mathbf{F}_n(\mathbf{U}_n) - \partial_y \mathbf{G}_n(\mathbf{U}_n) - \partial_z \mathbf{H}_n(\mathbf{U}_n)$ and $\mathbf{L}_i(\mathbf{U}_i) \equiv -\partial_x \mathbf{F}_i(\mathbf{U}_i) - \partial_y \mathbf{G}_i(\mathbf{U}_i) - \partial_z \mathbf{H}_i(\mathbf{U}_i)$.

An examination of eqs. (7) and (8) shows that the friction term in (8) can become very large relative to the rest of the force terms that act on the ions. This necessitates an implicit or semi-implicit treatment. Representing all the waves that arise in a two-fluid approximation also requires us to accurately preserve the relative magnitudes of the source terms $\gamma \rho_n \rho_i (\mathbf{v}_i - \mathbf{v}_n)$ and the rest of the flux terms $\mathbf{L}_i(\mathbf{U}_i)$ in eqn. (8). It is for that reason that we desire a formulation that has the following attributes: First, it should be genuinely second order accurate in time. Second, it should at least be semi-implicit and stable in the treatment of source terms. Third, it should permit us to obtain as accurate a treatment of the fluxes used in computing $\mathbf{L}_i(\mathbf{U}_i)$ and $\mathbf{L}_n(\mathbf{U}_n)$ as possible. To make it easy to implement the scheme we also require the fluxes in $\mathbf{L}_i(\mathbf{U}_i)$ and $\mathbf{L}_n(\mathbf{U}_n)$ to be evaluated using a basic time-explicit scheme. This calls for an implicit-explicit (IMEX) formulation which we develop below.

We formally describe the implicit-explicit scheme below. Let $\mathbf{U}_n^{(n)}$ and $\mathbf{U}_i^{(n)}$ represent the neutrals and ions at a time $t^n$ and let $\mathbf{U}_n^{(n+1)}$ and $\mathbf{U}_i^{(n+1)}$ denote the same quantities at a later time $t^{n+1} = t^n + \Delta t$. The predictor step of the IMEX scheme is:

$$\mathbf{U}'_n = \mathbf{U}_n^{(n)} + \frac{\Delta t}{2} \mathbf{L}_n(\mathbf{U}_n^{(n)}) - \frac{\Delta t}{2} \mathbf{S}(\mathbf{U}'_n, \mathbf{U}'_i) \tag{9}$$

$$\mathbf{U}'_i = \mathbf{U}_i^{(n)} + \frac{\Delta t}{2} \mathbf{L}_i(\mathbf{U}_i^{(n)}) + \frac{\Delta t}{2} \mathbf{S}(\mathbf{U}'_n, \mathbf{U}'_i) \tag{10}$$

The second order accurate corrector step is described by:

$$\mathbf{U}_n^{(n+1)} = \mathbf{U}_n^{(n)} + \Delta t\, \mathbf{L}_n(\mathbf{U}'_n) - \frac{\Delta t}{2}\left[\mathbf{S}(\mathbf{U}_n^{(n+1)}, \mathbf{U}_i^{(n+1)}) + \mathbf{S}(\mathbf{U}_n^{(n)}, \mathbf{U}_i^{(n)})\right] \tag{11}$$

$$\mathbf{U}_i^{(n+1)} = \mathbf{U}_i^{(n)} + \Delta t\, \mathbf{L}_i(\mathbf{U}'_i) + \frac{\Delta t}{2}\left[\mathbf{S}(\mathbf{U}_n^{(n+1)}, \mathbf{U}_i^{(n+1)}) + \mathbf{S}(\mathbf{U}_n^{(n)}, \mathbf{U}_i^{(n)})\right] \tag{12}$$

We explicitly write out our predictor stage for eqns. (9) and (10) by denoting the intermediate time variables with a prime. We also define two three-dimensional vectors $\mathbf{f}_n$ and $\mathbf{f}_i$ whose three components are given by the second, third and fourth components of $\mathbf{L}_n(\mathbf{U}_n^{(n)})$ and $\mathbf{L}_i(\mathbf{U}_i^{(n)})$. This leads to

$$\rho'_n \mathbf{v}'_n - \rho_n \mathbf{v}_n = \mathbf{f}_n \frac{\Delta t}{2} - \gamma \rho'_n \rho'_i (\mathbf{v}'_n - \mathbf{v}'_i)\frac{\Delta t}{2} \tag{13}$$

$$\rho'_i \mathbf{v}'_i - \rho_i \mathbf{v}_i = \mathbf{f}_i \frac{\Delta t}{2} - \gamma \rho'_n \rho'_i (\mathbf{v}'_i - \mathbf{v}'_n)\frac{\Delta t}{2} \tag{14}$$

Equations (13) and (14) are equivalent to the implicit scheme used by Tóth (1995) for a time step of $\Delta t/2$. Exclusive reliance on eqns. (13) and (14) would also make the scheme temporally first order accurate. We can solve for $\mathbf{v}'_n$ and $\mathbf{v}'_i$ to get the time-update for the predictor stage:

$$\mathbf{v}'_n = \frac{1}{\Delta}\left[\left(1+\gamma\rho'_n\frac{\Delta t}{2}\right)\left(\frac{\rho_n}{\rho'_n}\mathbf{v}_n + \frac{\Delta t}{2\rho'_n}\mathbf{f}_n\right) + \gamma\rho'_i\frac{\Delta t}{2}\left(\frac{\rho_i}{\rho'_i}\mathbf{v}_i + \frac{\Delta t}{2\rho'_i}\mathbf{f}_i\right)\right] \tag{15}$$

$$\mathbf{v}'_i = \frac{1}{\Delta}\left[\left(1+\gamma\rho'_i\frac{\Delta t}{2}\right)\left(\frac{\rho_i}{\rho'_i}\mathbf{v}_i + \frac{\Delta t}{2\rho'_i}\mathbf{f}_i\right) + \gamma\rho'_n\frac{\Delta t}{2}\left(\frac{\rho_n}{\rho'_n}\mathbf{v}_n + \frac{\Delta t}{2\rho'_n}\mathbf{f}_n\right)\right] \tag{16}$$

where

$$\Delta = 1 + \gamma(\rho'_{n+} + \rho'_i)\frac{\Delta t}{2} \tag{17}$$

We formulate our Runge-Kutta corrector step for equations (11) and (12) in a half-implicit manner, so that we have second-order temporal accuracy. We also define two three-dimensional vectors $\mathbf{f}'_n$ and $\mathbf{f}'_i$ whose three components are given by the second, third and fourth components of $\mathbf{L}_n(\mathbf{U}'_n)$ and $\mathbf{L}_i(\mathbf{U}'_i)$.

$$\rho_n^{(n+1)}\mathbf{v}_n^{(n+1)} - \rho_n\mathbf{v}_n = \mathbf{f}'_n\Delta t - \gamma\rho_n^{(n+1)}\rho_i^{(n+1)}\left(\mathbf{v}_n^{(n+1)} - \mathbf{v}_i^{(n+1)}\right)\frac{\Delta t}{2} - \gamma\rho_n\rho_i(\mathbf{v}_n - \mathbf{v}_i)\frac{\Delta t}{2} \tag{18}$$

$$\rho_i^{(n+1)}\mathbf{v}_i^{(n+1)} - \rho_i\mathbf{v}_i = \mathbf{f}'_i\Delta t - \gamma\rho_n^{(n+1)}\rho_i^{(n+1)}\left(\mathbf{v}_i^{(n+1)} - \mathbf{v}_n^{(n+1)}\right)\frac{\Delta t}{2} - \gamma\rho_n\rho_i(\mathbf{v}_i - \mathbf{v}_n)\frac{\Delta t}{2} \tag{19}$$

Equations (18) and (19) have the solution

$$\mathbf{v}_n^{(n+1)} = \frac{1}{\Delta}\left[\left(1+\gamma\rho_n^{(n+1)}\frac{\Delta t}{2}\right)\Psi_n + \gamma\rho_i^{(n+1)}\frac{\Delta t}{2}\Psi_i\right] \tag{20}$$

$$\mathbf{v}_i^{(n+1)} = \frac{1}{\Delta}\left[\left(1+\gamma\rho_i^{(n+1)}\frac{\Delta t}{2}\right)\Psi_i + \gamma\rho_n^{(n+1)}\frac{\Delta t}{2}\Psi_n\right] \tag{21}$$

where we use the two auxiliary vectors

$$\Psi_n = \frac{1}{\rho_n^{(n+1)}}\left[\rho_n\mathbf{v}_n + \mathbf{f}'_n\Delta t - \gamma\rho_n\rho_i\frac{\Delta t}{2}(\mathbf{v}_n - \mathbf{v}_i)\right] \tag{22}$$

$$\Psi_i = \frac{1}{\rho_i^{(n+1)}}\left[\rho_i\mathbf{v}_i + \mathbf{f}'_i\Delta t - \gamma\rho_n\rho_i\frac{\Delta t}{2}(\mathbf{v}_i - \mathbf{v}_n)\right] \tag{23}$$

The half-implicit corrector step requires a time step constraint in order to maintain stability. We require that time step $\Delta t$ is bounded by

$$\Delta t \leq \frac{1}{2}\frac{\sqrt{c_s^2 + v'^2_A}}{\gamma\rho'_n|\mathbf{v}'_n - \mathbf{v}'_i|} \tag{22}$$

where $v'_A$ is the Alfvén speed at the intermediate time step. This is equivalent to stating that the maximum change in the ionized fluid velocity that we permit in one time step is bounded by the fast magnetosonic speed.

We have found that a conservative strategy for the fluxes from the self-gravity was necessary to correctly track the propagation of self-gravitating waves. We first use the Poisson equation to re-write the gravitational force term in the momentum equation as

$$\mathbf{f}_g = -\rho \mathbf{g} = -\frac{\mathbf{g}}{4\pi G} \nabla \cdot (\mathbf{g}) \qquad (23)$$

We then calculate the forces on the zone centres. For example, in the x-direction Equation (23) becomes

$$f_x = -\frac{1}{4\pi G}\left[\frac{1}{2}\frac{\partial}{\partial x}\left(g_x^2 - g_y^2 - g_z^2\right) + \frac{\partial}{\partial y}\left(g_x g_y\right) + \frac{\partial}{\partial z}\left(g_x g_z\right)\right] \qquad (24)$$

where we have used the fact that $\nabla \times \mathbf{g} = 0$ to subtract $\mathbf{g} \times (\nabla \times \mathbf{g})$ from the right-hand side of Equation 24.

## 3. Tests of Algorithm

We utilize four test problems in order to verify the results produced by the algorithm presented in Section 2. In Subsection 3.1 we numerically reproduce the dispersion analysis of self-gravitating MHD waves of Balsara (1996), and demonstrate that we can capture the predicted decay rates for fast magnetosonic and Alfvén waves, and the propagation of slow waves. In Subsection 3.2 we set up a C-shock (Draine 1980, Wardle 1990) and demonstrate that we can follow its evolution (e.g. Tóth 1994). In Subsection 3.3, we present a test of a circular blast wave that allows us to examine the behaviour of the code at arbitrary angles to the magnetic field. In Subsection 3.4, we numerically reproduce the instability found by Wardle (1990) for a C-shock.

### 3.1 Eigenvalue analysis of MHD self-gravitating waves

Balsara (1996) has considered the propagation of MHD waves in a partially ionized, self-gravitating molecular cloud. He found the fast magnetosonic waves and Alfvén waves were strongly damped on wavelengths shorter than the damping scale, while slow magnetosonic waves could continue to propagate with only slight damping. Furthermore, on scales larger than the Jeans length, the slow mode couples with self-gravity to modulate gravitational collapse, while fast and Alfvén waves remain slightly damped. The ability of the algorithm presented in Section 2 to reproduce the damping rates on both sides of the damping scale, as well as on scales above the Jeans length, ensures that the algorithm will function as intended when wave propagation on different scales are important.

We set up our computational domain on a one-dimensional grid with periodic boundary conditions. We consider runs at ionization fractions of $\xi=10^{-2}$, $10^{-4}$, $10^{-5}$, and $10^{-6}$. We set our neutral density such that the gravitational frequency defined in Balsara (1996), $\omega_J = \sqrt{4\pi G\rho_{0n}} = 9.46\times 10^{-6}$ yr$^{-1}$, where $\rho_{0n}$ is the unperturbed density of the neutral fluid; this corresponds to a density of $\rho_{0n} = 1.07\times 10^{-19}$ g cm$^{-3}$. We set the isothermal sound speed in both fluids to $c_s = 0.2$ km s$^{-1}$, and a magnetic field of $\mathbf{B} = (46.4\hat{\mathbf{x}} + 23.2\hat{\mathbf{y}})\,\mu$G, corresponding to an Alfvén speed of $\mathbf{v}_A = 0.447$ km s$^{-1}$. We perturb these constant states with the eigenvectors of the system (Balsara 1996), with an amplitude of $10^{-5}$. We use resolutions of 128 zones per wavelength and compare the results with that from an otherwise identical run with just 32 zones per wavelength. We use perturbations with wavenumbers (normalized to the dissipation wavelength $\tilde{L} = (B_{0x}/\gamma\rho_{0i})(4\pi\rho_{0n})^{-1/2} = 0.568(\xi/10^{-4})^{-1}$ AU) of $\tilde{k}=10^{-5}$, $10^{-4}$, $10^{-3}$, $10^{-2}$, $10^{-1}$, 1.0, 10. As a result, the length of our computational domain per wavelength is $2\pi\tilde{L}/\tilde{k}=(1.73\times 10^{-5}$ pc) $(\xi/10^{-4})^{-1}\tilde{k}^{-1}$, corresponding to scales of $1.83\times 10^{-6}$ pc – 1.73 pc for ionization fractions of $10^{-4}$ for the different values of $\tilde{k}$ above. At ionization fractions of $10^{-4}$, $10^{-5}$ and $10^{-6}$, the slow mode at some of these wavelengths is Jeans unstable, and is thus expected to grow. (The Jeans length is $L_J$=0.136 pc in this system for all ionization fractions, and corresponds to $\tilde{k}_J=1.267\times 10^{-8}\,\xi^{-1}$.) We present a summary of our runs for this test problem in Table 1. To facilitate the use of this test problem in future numerical studies, we present in Appendix A the eigenvectors of the slow magnetosonic waves that we used for the runs in Table 1.

We compare the measured growth and decay rates in the neutral density of each simulation to the values predicted from the eigenvalue analysis in Fig. 1, and the growth and decay rates of the ionized fluid density in Fig. 2. We have run the dispersion analysis using two forms of limiters – MinMod (diamonds) and WENO (stars). The WENO schemes are described in Jiang & Shu (1996) and Balsara & Shu (2000), and the specific implementation is given in Balsara (2004). The lines show the predicted imaginary component of the eigenvalue, with positive values (growing modes) shown in the upper portion of the plot on a logarithmic scale, and negative values (decaying modes) shown on the lower portion, again on a logarithmic scale. In general, the modes at these wavenumbers are decaying (the imaginary component of the eigenvalue is negative), but at long wavelengths (small wavenumbers) the slow mode becomes gravitationally unstable and grows. We find that we generally get good agreement with the predicted eigenvalues with the MinMod limiters when the decay rate $|\omega_{\text{imag}}|/k \geq 10^{-4}$ km s$^{-1}$. When we use the higher-order WENO limiter, however, we see significant improvement in our ability to capture very small decay rates, on the order of $10^{-6}$ km s$^{-1}$, an improvement of two orders of magnitude.

For practical work, achieving such high resolutions is likely not possible. We present the growth and decay rates of the same simulations, but run with a resolution of 32 zones per wavelength of the mode, in Figs. 3 (from the neutral fluid) and 4 (from the ionized fluid). At these more modest resolutions, we see that we still get good agreement with the WENO limiters for decay rates greater than $|\omega_{\text{imag}}|/k \geq 10^{-4}$ km s$^{-1}$. When we use the MinMod limiter, however, we find that in the presence of large coupling between the

fluids, the decay rates of the modes are greatly enhanced. With weaker coupling between the fluids, we find that some of the wavelengths predicted to be decaying grow instead. Figs. 4c and 4d actually correspond to ionization fractions that are observed in star-forming systems. It is important to observe from Figs. 4c and 4d that the use of the MinMod limiter would produce a specious gravitational collapse in the ions even when the analytic solution does not predict such growth.

We compare the phase speed of the simulation data to the predicted real part of the slow mode eigenvalue from the dispersion analysis in Figs. 5 (from the neutral fluid) and 6 (from the ionized fluid), again on the low-resolution mesh of 32 zones per wavelength. We see that the WENO limiter again does very well in matching the predicted phase speed. The MinMod limiter performs adequately at ionization fractions of $10^{-2}$ and $10^{-4}$, where the coupling between the fluids is strong, but does poorly at lower ionization fractions.

We see, therefore, that highly accurate schemes are necessary to properly capture the behaviour of MHD waves, even at modest ionization fractions and on large spatial scales near the Jeans length. This is especially important on resolution-starved grids, where one can afford just a few zones per mode.

We also perform a convergence study of the ability of the code to reproduce the eigenvalues. For this test, we initialized the eigenvectors of an Alfvén wave at a wavelength of $\tilde{k}=1.0$ at an ionization fraction of $10^{-4}$. We examined the L1 errors of the decay rate of the z-component of the magnetic field for resolutions of 64, 96, 128, 196, and 256 zones per wavelength. The convergence is shown in Table 2, with an asymptotic slope of -1.82 indicating that we are indeed achieving second-order accuracy as we expected.

**3.2 Ion-Neutral C-shock**

C-shocks, or continuous shocks, can develop in partially ionized flows when the shock speed is supersonic, but not super-Alfvénic in the ionized fluid (Draine 1980). The structure of C-shocks has been studied extensively in the literature (Wardle 1990; Tóth 1994; Smith & Mac Low 1997), and in multiple dimensions is subject to an instability (Wardle 1990; Tóth 1994; Mac Low & Smith 1997, Ciolek & Roberge 2002, Falle 2003).

We draw our initial conditions from the similar test in Tóth (1994). We use a temperature in both fluids of 20 K, a magnetic field parallel to the shock with field strength of 5 μG, a density in the neutral fluid of $2.338 \times 10^{-22}$ g cm$^{-3}$, and a density in the ionized fluid of $5.010 \times 10^{-25}$ g cm$^{-3}$. We use an isothermal equation of state, with a sound speed of 0.344 km s$^{-1}$. We use a one-dimensional mesh of 256 zones, with the total size of the grid being 0.02 pc. The fluid is given a velocity of -2.2605 km s$^{-1}$ and is driven into a reflecting wall at the origin, simulating a 3.0 km s$^{-1}$ piston-driven shock.

We show the density structure of the shock in both the ionized and neutral fluids that develops at a time of $1.6 \times 10^4$ years in Fig. 7a, along with the reference solutions.

Fig. 7b shows the error in the numerical solution. We reproduce to the predicted solution to within 2 percent in the neutral fluid, and better than 1 percent in the ionized fluid.

### 3.3 Blast Wave Problem in a Partially Ionized Plasma

A third test that we perform is that of a circular blast wave that develops into a multi-dimensional shock. This allows us to test the ability of the code to handle shocks that are oblique to the magnetic field.

We set up our pulse on a two-dimensional Cartesian grid of with dimension 0.2 pc on a side, with a density $\rho_n$=9.95x10$^{-21}$ g cm$^{-3}$ in the neutral fluid and ionization fractions of $\xi$=10$^{-6}$, 10$^{-4}$, and 10$^{-2}$. We use an isothermal sound speed of 0.2 km s$^{-1}$ in both fluids. We placed a uniform magnetic field oriented in the y-direction with a field strength of 10 µG, leading to an Alfvén number for the neutral fluid, $A_n^2 = 8\pi\rho_n c_s^2/B^2 = 1$. The Alfvén speed in the ions will thus be a factor of $(m_i\xi/m_n)^{-1/2}$ faster. We initialize a circular pulse of radius 0.02 pc with a density ten times larger than the ambient in both the neutral and ionized fluids. We use a grid of $256^2$ zones.

The resulting density and x-velocity for each fluid are shown in Fig. 8 for each of the ionization fractions we studied, at a time of t=0.1. We can see that there is a clear transition, from an ionization fraction of 10$^{-2}$ where the ionized and neutral fluids track each other very well, to an ionization fraction of 10$^{-6}$ where the neutral fluid is almost completely unaffected by the presence of the magnetized ionized fluid. At an ionization fraction of 10$^{-4}$, we see an intermediate result, with the neutral fluid picking up the magnetohydrodynamic shock in the velocity structure, but showing a weak sign of it in the density.

We can see this more clearly by looking at a slice through the midplane of the blast wave. We show this in Fig. 9 for an ionization fraction of 10$^{-2}$, in Fig. 10 for an ionization fraction of 10$^{-4}$, and in Fig. 11 for an ionization fraction of 10$^{-6}$. Fig. 9a shows clearly that the density structures of both the neutral and the ionized components for an ionization fraction of 10$^{-2}$ track each other very well, due to the large coupling between the two species. Fig 9b shows that the velocity structures in the two fluids track each other even more closely, as the two lines overlap everywhere. At an ionization fraction of 10$^{-4}$, we see that where the ionized fluid has a shock, the neutral fluid is largely continuous. The velocities of the two fluids, however, still track each other quite well, even with the reduced coupling between the two species. At an ionization fraction of 10$^{-6}$, we see that the coupling between the two fluids is so weak that the neutral fluid is largely unaffected by the magnetized fluid. (It is worth pointing out that a magnetized precursor has traveled off the grid at the time of the image, moving at the Alfvén speed that is much larger than the sound speeds in the problem.)

We contrast the above results with slices of density and y-velocity through the vertical midplane at the same times as those in Figs. 9, 10, and 11. We show these in Fig.

12 for an ionization fraction of $10^{-2}$, Fig. 13 for an ionization fraction of $10^{-4}$, and Fig. 14 for an ionization fraction of $10^{-6}$.  In this case, as the magnetic field is completely perpendicular to the shock front, the magnetic field is both continuous across the shock and does not contribute to the momentum equation of the ionized fluid.  As a result, the ions evolve in the same manner as the neutrals, and we see that in Figs. 12b, 13b, and 14b that the evolution of the magnetized shock in this direction does not depend on the ionization fraction at all, and is identical to the velocity structure in the neutral fluid.  We do see a difference in the density jump in the shock in the ionized densities, as the x-component of the magnetic field does grow due to the curvature in the shock.

### 3.4 Wardle Instability of a C-shock

As a final test problem, we consider the instability studied by Wardle (1990) of a C-shock.  We draw our initial conditions from the similar test in Wardle (1990) and Tóth (1994).  We use a temperature in both fluids of 20 K, a magnetic field parallel to the shock with field strength of 5 µG, a density in the neutral fluid of $2.338 \times 10^{-22}$ g cm$^{-3}$, and a density in the ionized fluid of $5.010 \times 10^{-25}$ g cm$^{-3}$.  We give the fluid a velocity of -12.5 km s$^{-1}$, and initialize the C-shock solution for these parameters.  The shock thickness for this test case is $L_{shk}=7.0 \times 10^{15}$ cm (Wardle 1990), and the critical wavelength for this shock is 0.71 $L_{shk}$ (Tóth 1994).  We initialize the solution on a two-dimensional grid that is (4 $L_{shk}$, 0.71 $L_{shk}$) and (256, 64) zones in size.  We use the perturbing solution of Tóth (1994) (his Equation 4.4) to give a sinusoidal perturbation to the y-velocities of both the ionized and neutral fluids.  We use periodic boundary conditions in y, continuative boundary conditions at the lower x boundary (downstream from the shock), and an inflow boundary condition at the upper x boundary (upstream from the shock).

We show the logarithm of the ion density at three different times in Fig. 15: t=0, t=0.9 kyr, and t=1.5 kyr, by which time the instability has fully set in.  The growth of the instability, as measured by the standard deviation of $B_x$ (Tóth 1994), is plotted in Fig. 16.  We clearly see a linear regime where the perturbation grows after it encounters the C-shock.  We measure the growth rate in the linear regime to be $1.83 \times 10^{-10}$ s$^{-1}$ = 9.87 $t_{flow}^{-1}$, where $t_{flow}=5.4 \times 10^{10}$ s (Wardle 1990).  The value found by Wardle (1990) was 9.10 $t_{flow}^{-1}$. The difference between the analytical and numerical solution is about 8%. Part of the difference can be accounted for by numerical error but part of it could also be due to the specific nature of the problem. Numerical methods respond best when there is a continuous train of sinusoids interacting with the shock whereas the present test problem uses a single sinusoidal pulse. Even so, the reasonably good agreement that we obtain on rather small meshes is satisfying.

### 4. Discussion and Conclusions

We have presented a semi-implicit predictor-corrector strategy for two-fluid ion-neutral friction that is second-order accurate in space and time.  The method has been implemented in the RIEMANN code for astrophysical fluid dynamics.  We have showed that the method can capture the propagation of MHD waves accurately at scales both above and below the dissipation length, as well as at scales above and below the Jeans length. This wave analysis demonstrated the need for highly accurate slope limiters to reproduce the propagation of these waves. In particular, low-order limiters can greatly overestimate the decay of MHD waves, effectively increasing the dissipation rate of kinetic energy. Furthermore, low-order limiters are more susceptible to allowing wave modes to grow instead of decay, particularly when the decay rates are very small. These effects are exacerbated at very low ionization fractions. This latter problem may be alleviated by using the heavy-ion approximation (Li, McKee & Klein 2006; Oishi & Mac Low 2006; Li et al. 2007), which involves raising the ionization fraction to levels that do not cause numerical difficulties while correspondingly decreasing the ion-neutral drag coefficient to maintain the ambipolar drift forces at a constant level. This method was proposed by Li, McKee & Klein (2006) and Oishi & Mac Low (2006) as a means to reduce the constraint on the time step imposed by the Alfvén speed in the ionized fluid. However, we showed in Figs. 3 and 4 that a low-order limiter still has difficulties in capturing the decay rate of the waves, even at large ionization fractions, unless one can afford a very large number of zones for the simulation. Thus, the incorporation of higher-order methods is desirable even under the heavy-ion approximation.

We have shown that our numerical method also works robustly for capturing the behaviour of C-shocks. We are able to maintain the structure of a C-shock for long periods of time. We show that we can follow the evolution of a circular C-shock as it expands at all angles oblique to the magnetic field. We are also able to match the growth rate of a sinusoidal perturbation impacting on a C-shock in multiple dimensions.

**Acknowledgements**

DSB acknowledges support via NSF grant AST-0607731. DSB also acknowledges NASA grants HST-AR-10934.01-A, NASA-NNX07AG93G and NASA-NNX08AG69G. The majority of simulations were performed on PC clusters at UND but a few initial simulations were also performed at NASA-NCCS.

**Appendix A: Eigenvectors for Two-Fluid MHD Dispersion Analysis**

We reproduce in this Appendix the eigenvectors and eigenvalues used for the dispersion analysis in Section 3.1. We utilize the nomenclature of Balsara (1996) and the characteristic matrix given in the appendix of that reference.

We solve the eigensystem using the LAPACK algorithms that are implemented in IDL. The eigenvectors produced by these routines are normalized to a unit length, and rotated such that the largest perturbation has no imaginary component. We further multiply these perturbations by an amplitude of $10^{-5}$ to ensure that we remain in the linear regime as we evolve the waves. For the dispersion study, we examine only the slow modes. For the resolution study, we utilize the Alfvén mode.

Table 3 gives the eigenvectors for each wavenumber for Run A, at an ionization fraction of $10^{-2}$. Table 4 gives the eigenvectors for each wavenumber for Run B, with an ionization fraction $10^{-4}$. Table 5 gives the eigenvectors for each wavenumber for Run C, with an ionization fraction of $10^{-5}$. Table 6 gives the eigenvectors for each wavenumber for Run D, with an ionization fraction of $10^{-6}$. Tables 3-6 also list the eigenvalues of the slow mode at each wavenumber, expressed as $\omega/k$ (as in Balsara 1996) with units of velocity. The real part of the eigenvalue as listed in the tables is thus the phase speed of the wave. We express the wavenumbers in Tables 3-6 as the dimensionless wavenumber $\tilde{k} = k\tilde{L}$, where $\tilde{L}$ is the damping length $\tilde{L} = b_x/\gamma\rho_{0i}$, such that $\tilde{k} = k\tilde{L}$.

The resolution study in Section 3.1 used an ionization fraction of $10^{-4}$ and a wavenumber $\tilde{k}=1.0$. For this test, we used the eigenvectors corresponding to an Alfvén wave. The eigenvalues for this wave, expressed as a phase velocity, is $\omega/k = (34655.5 - 19974.8i)$ cm s$^{-1}$. The eigenvectors for each fluid variable are thus

$$\tilde{\rho}_{1n} = (1.6900054 + 2.0419427i) \times 10^{-20} \quad \tilde{\rho}_{1i} = (0.49512651 + 2.0971125i) \times 10^{-20}$$
$$v_{1nx} = (9.9355355 + 3.7007198i) \times 10^{-16} \quad v_{1ix} = (6.1383582 - 5.0653335i) \times 10^{-16}$$
$$v_{1ny} = (-1.6450722 - 0.35644265i) \times 10^{-15} \quad v_{1iy} = (0.19396856 + 2.4128802i) \times 10^{-15}$$
$$v_{1nz} = (-4.9968495 - 2.8873543i) \times 10^{-1} \quad v_{1iz} = (-5.0031498 + 2.8837138i) \times 10^{-1}$$
$$b_{1y} = (1.7191396 - 0.68110079i) \times 10^{-15} \quad b_{1z} = 5.7747179 \times 10^{-1}$$

Table 1: Simulations for Two-Fluid MHD Dispersion Analysis

| Run | Ionization Fraction | Dissipation Length $\tilde{L}$ (cm) |
|---|---|---|
| A | $10^{-2}$ | $8.471 \times 10^{10}$ |
| B | $10^{-4}$ | $8.471 \times 10^{12}$ |
| C | $10^{-5}$ | $8.471 \times 10^{13}$ |
| D | $10^{-6}$ | $8.471 \times 10^{14}$ |

Table 2: Convergence Tests. The predicted ratio of the decay rate to the wavenumber from the eigenvalue analysis is -19974.751 cm s$^{-1}$.

| Resolution | Im($\omega$)/k (cm s$^{-1}$) | L1 Error | Slope |
|---|---|---|---|
| 64 | -20122.904 | 0.007416 | -- |
| 96 | -20028.500 | 0.002691 | -2.501 |
| 128 | -20004.332 | 0.001481 | -2.076 |
| 196 | -19987.943 | 0.000660 | -1.992 |
| 256 | -19983.370 | 0.000431 | -1.909 |

Table 3: Eigenvectors for Run A, the slow mode for an ionization fraction of $10^{-2}$. The eigenvalues for these modes are expressed as phase velocities, for each of the normalized wavenumbers $\tilde{k}$. For these simulations, $\tilde{L}=8.471 \times 10^{10}$ cm = $5.683 \times 10^{-3}$ AU.

| | 10 | 1 | 0.1 | 0.01 | 0.001 | $10^{-4}$ | $10^{-5}$ |
|---|---|---|---|---|---|---|---|
| $\omega/k$ | 19757.5 -407.6i | 17891.2 -1052.7i | 17426.7 -121.1i | 17421.6 -12.1i | 17421.6 -1.2 i | 17420.1 -0.1 i | 17272.9-0.0118ι |
| $\tilde{\rho}_{1n}$ | 3.7409885e-05 + 7.7186290e-07 i | 3.4324136e-05 + 2.0195393e-06 i | 3.3679774e-05 + 2.3401987e-07 i | 3.3673063e-05 + 2.3433043e-08 i | 3.3673031e-05 + 2.3433344e-09 i | 3.3676432e-05 + 2.3432253e-10 i | 3.4021061e-05 + 2.3321358e-11 i |
| $v_{1nx}$ | 0.73943983 | 0.61622615 | 0.58695095 | 0.58663951 | 0.58663649 | 0.58664645 | 0.58764254 |
| $v_{1ny}$ | -0.0011167742 + 0.061586514 i | 0.27856938 + 0.19690815 i | 0.37191775 + 0.023874142 i | 0.37298361 + 0.0023918045 i | 0.37299417 + 0.00023918451 i | 0.37298362 + 2.3915079e-05 i | 0.37191732 + 2.3577757e-06 i |
| $v_{1nz}$ | 3.5423075e-16 - 1.0933512e-16 i | -2.7067790e-16 + 2.9016077e-16 i | 4.7377697e-16 + 1.2236696e-15 i | -5.4836787e-16 + 5.4397598e-14 i | -6.7115577e-14 + 4.9981638e-13 i | -1.5299516e-15 + 5.7455574e-12 i | 1.1795546e-14 + 9.9564226e-11 i |
| $\tilde{\rho}_{1i}$ | 2.9599837e-05 + 5.0961574e-06 i | 3.2074347e-05 + 5.6664334e-06 i | 3.3652917e-05 + 6.9912496e-07 i | 3.3672794e-05 + 7.0056203e-08 i | 3.3673028e-05 + 7.0057620e-09 i | 3.3676432e-05 + 7.0055210e-10 i | 3.4021061e-05 + 6.9810906e-11 i |
| $v_{1ix}$ | 0.58689573 + 0.088620928 i | 0.57981368 + 0.067615623 i | 0.58653925 + 0.0081084338 i | 0.58663538 + 0.00081225410 i | 0.58663645 + 8.1226735e-05 i | 0.58664645 + 8.1217597e-06 i | 0.58764254 + 8.0301155e-07 i |
| $v_{1iy}$ | 0.30319567 + 0.060826269 i | 0.35931145 + 0.067127606 i | 0.37284528 + 0.0076637972 i | 0.37299290 + 0.00076730283 i | 0.37299427 + 7.6731053e-05 i | 0.37298362 + 7.6715600e-06 i | 0.37191732 + 7.5175261e-07 i |
| $v_{1iz}$ | -3.8028734e-18 | -7.1358264e-16 | 4.1109836e- | - | - | - | 1.172113 |
| | - 1.6057719e-17 i | + 4.9727124e-16 i | 16 + 1.1625561e-15 i | 5.5628229e-16 + 5.4385401e-14 i | 6.5887018e-14 + 4.9996397e-13 i | 1.5800636e-15 + 5.7454368e-12 i | 6e-14 + 9.9564211e-11 i |
| $b_{1y}$ | -0.019036914 - 0.033829844 i | -0.15026708 - 0.083335542 i | -0.18258734 - 0.0095539250 i | -0.18293346 - 0.00095656207 i | -0.18293670 - 9.5657161e-05 i | -0.18291615 - 9.5635837e-06 i | -0.18085216 - 9.3505942e-07 i |
| $b_{1z}$ | 7.0253558e-18 + 3.2654595e-17 i | 3.1652462e-16 + 3.2516428e-16 i | -1.1425582e-15 - 7.8112642e-17 i | 4.8939895e-16 + 3.6330437e-18 i | 1.5166706e-13 - 1.1479509e-12 i | 2.6401978e-15 + 2.0101648e-19 i | -5.8289334e-16 - 1.7995853e-20 i |

Table 4: Eigenvectors for Run B, the slow mode for an ionization fraction of $10^{-4}$. The eigenvalues for these modes are expressed as phase velocities, for each of the normalized wavenumbers $\tilde{k}$. For these simulations, $\tilde{L}$=8.471x$10^{12}$ cm = 0.5683 AU.

| | 10 | 1 | 0.1 | 0.01 | 0.001 | $10^{-4}$ | $10^{-5}$ |
|---|---|---|---|---|---|---|---|
| $\omega/k$ | 19946.2 - 394.8 i | 18005.0 -1134.1i | 17485.6 - 131.67 i | 17478.8- 13.182 i | 17346.3 - 1.292 i | -2.67e-6+14104.1 i | -3.832e-7 +252042.1i |
| $\tilde{\rho}_{1n}$ | 3.7186528e-5 +7.3599208e-7i | 3.4548792e-05 + 2.1762436e-06i | 3.3967558e-05 + 2.5578169e-07i | 3.3964560e-05 + 2.5614353e-08i | 3.4266302e-05 + 2.5514607e-09i | -7.5472880e-15 - 4.5576892e-05i | -3.3126204e-18 - 2.8011081e-06i |
| $v_{1nx}$ | 0.74202227 | 0.62451879 | 0.59397837 | 0.59365955 | 0.59439414 | 0.64282006 | 0.70599716 |
| $v_{1ny}$ | 0.0058835998 + 0.059044695 i | 0.27375894 + 0.19280826 | 0.36593999 + 0.023429927 i | 0.36697941 + 0.0023469915 | 0.36613973 + 0.00023182067i | 0.28581747 - 4.9063999e-11i | 0.0086618185 + 3.6274371e-12i |
| $v_{1nz}$ | -2.8226224e-17 - 2.8317443e-16i | 2.2848420e-16 - 4.1912571e-16i | 4.1217530e-18 - 2.3496401e-16i | 7.5148094e-17 - 5.0519526e-15i | 1.9904183e-17 + 2.9648186e-14i | -1.2147021e-12 + 9.9839414e-15i | -5.9456489e-15 + 2.3226166e-14i |
| $\tilde{\rho}_{1i}$ | 2.9809663e-05 + 1.5181236e-06i | 3.2140968e-05 + 5.6000593e-06i | 3.3938303e-05 + 7.1320904e-07i | 3.3964267e-05 + 7.1486780e-08i | 3.4266299e-05 + 7.1282075e-09i | -1.2568784e-14 - 4.5576535e-05i | 1.7630249e-15 - 2.8011071e-06i |
| $v_{1ix}$ | 0.59519026 + 0.018512785 i | 0.58504903 + 0.064376579 i | 0.59352707 + 0.0080022647i | 0.59365504 + 0.00080179812i | 0.59439410 + 7.9389706e-05i | 0.64281502 - 1.5674328e-15i | 0.70599689 - 3.7077851e-16i |
| $v_{1iy}$ | 0.29973296 + | 0.35278466 + | 0.36684375 + | 0.36698845 + | 0.36613982 + | 0.28582755 - | 0.086623643 |

| | 0.023878430i | 0.064115794i | 0.0074254714i | 0.00074339538i | 7.3041261e-05i | 4.9071415e-11i | + 3.6249822e-12i |
|---|---|---|---|---|---|---|---|
| $v_{1iz}$ | -1.4160467e-17 + 9.8038113e-19i | 7.7695387e-17 - 2.8202190e-16i | 2.9726011e-17 - 2.3045036e-16i | 5.7348762e-17 - 5.0530434e-15i | -1.8705716e-17 + 2.9648117e-14i | -1.2146454e-12 +9.9838317e-15i | -5.9456234e-15 + 2.3226167e-14i |
| $b_{1y}$ | -0.0037053778 - 0.029396219i | -0.12889498 - 0.079049502i | -0.16024678 - 0.0090401724i | -0.16056182 - 0.00090488655i | -0.15897961 - 8.8733236e-05i | 3.5298314e-11 - 0.10090688i | -6.3282780e-13 - 0.054647392i |
| $b_{1z}$ | 2.8425031e-17 - 1.4034588e-18i | 1.5288961e-16 + 1.2008071e-16i | 2.8435130e-17 + 2.0281851e-18i | -8.9466423e-17 - 6.4006943e-19i | -8.8836997e-17 - 6.2860833e-20i | -2.9845051e-19 - 1.6118886e-16i | 3.5039439e-22 - 3.3427001e-20i |

Table 5: Eigenvectors for Run C, the slow mode for an ionization fraction of 10$^{-5}$. The eigenvalues for these modes are expressed as phase velocities, for each of the normalized wavenumbers $\tilde{k}$. For these simulations, $\tilde{L}$=8.471x10$^{13}$ cm = 5.683 AU.

| | 10 | 1 | 0.1 | 0.01 | 0.001 | 10$^{-4}$ | 10$^{-5}$ |
|---|---|---|---|---|---|---|---|
| $\omega/k$ | 19948.071-394.84724i | 18006.064 - 1134.9676i | 17484.858 - 131.75218 i | 17347.032 - 12.926902 i | 5.9603096e-07+ 14082.779 i | -2.0015577e-06+ 251896.63 i | -0.00013122661+ 2534443.0 i |
| $\tilde{\rho}_{1n}$ | 3.7185142e-05+ 7.3603360e-07i | 3.4550671e-05+ 2.1778159e-06i | 3.3973021e-05+ 2.5599405e-07i | 3.4268467e-05+ 2.5536651e-08i | 2.0586124e-15 - 4.5646349e-05i | -1.6526073e-17 - 2.8027253e-06i | 6.5455464e-18 - 2.7899455e-07i |
| $v_{1nx}$ | 0.74206249 | 0.62459336 | 0.59404718 | 0.59445654 | 0.64282746 | 0.70599706 | 0.70709579 |
| $v_{1ny}$ | 0.0059472659+ 0.059015869i | 0.27371471+ 0.19277054i | 0.36587669+ 0.023422984i | 0.36607614+ 0.0023177999i | 0.28585135+ 8.5437762e-12i | 0.0086810240 - 1.1285488e-11i | 8.8031919e-05+ 4.4033982e-15i |
| $v_{1nz}$ | 1.0395712e-18 - 6.8729194e-18i | 5.1286096e-17+ 8.3904125e-17i | 6.3267502e-16 - 1.4701328e-15i | -4.7644986e-16 - 1.6415511e-14i | 9.1412399e-15+ 6.5269682e-18i | 1.8195928e-13 - 2.9798241e-13i | 8.3413665e-19+ 4.9561884e-16i |
| $\tilde{\rho}_{1i}$ | 2.9808536e-05+ 1.4841116e-06i | 3.2141267e-05+ 5.5994328e-06i | 3.3943745e-05+ 7.1334015e-07i | 3.4268177e-05+ 7.1296168e-08i | 2.0653534e-15 - 4.5642776e-05i | -2.2984232e-14 - 2.8027145e-06i | 1.1902904e-17 - 2.7899454e-07i |
| $v_{1ix}$ | 0.59520880 + 0.017835347i | 0.58509290 + 0.064344450i | 0.59359556 + 0.0080004888i | 0.59445211 + 0.00079379557i | 0.64277714 + 9.2352443e-15i | 0.70599433 - 5.3749086e-15i | 0.70709576 + 1.7139068e-15i |
| $v_{1iy}$ | 0.29967339 + | 0.35272425 + | 0.36678004 + | 0.36608501 + | 0.28595199 + | 0.0086864908 - | 8.8087697e-05+ |
| | 0.023531185i | 0.064087719i | 0.0074220132i | 0.00073020877i | 8.5273334e-12i | 1.1288010e-11i | 8.8734098e-15i |
| $v_{1iz}$ | -1.3962409e-16+ 4.0066923e-17i | 4.8678820e-18 - 2.0017274e-16i | 1.1773576e-16 - 1.6284097e-15i | 2.8933152e-16 - 1.6396977e-14i | 8.8063329e-15+ 6.5559191e-18i | 1.8195922e-13 - 2.9798241e-13i | 8.3496126e-19+ 4.9561885e-16i |
| $b_{1y}$ | -0.0035673409 - 0.029373717i | -0.12870307 - 0.079011911i | -0.16002992 - 0.0090338190i | -0.15877913 - 0.00088689299i | -1.7264754e-11 - 0.10065222i | -2.4615916e-12 - 0.054674915i | 2.2578484e-13 - 0.0055785005i |
| $b_{1z}$ | 1.3440408e-18+ 1.4462824e-17i | -1.6027648e-16 - 1.2874817e-16i | -1.1981040e-15 - 8.5713215e-17i | 1.3662560e-15+ 9.6681815e-18i | 2.4750177e-21+ 1.4460296e-15i | -3.7640951e-21 - 4.3075403e-20i | 5.8733852e-25+ 3.0493214e-20i |

Table 6: Eigenvectors for Run D, the slow mode for an ionization fraction of $10^{-6}$. The eigenvalues for these modes are expressed as phase velocities, for each of the normalized wavenumbers $\tilde{k}$. For these simulations, $\tilde{L}$=8.471x10$^{14}$ cm = 56.83 AU.

| | 10 | 1 | 0.1 | 0.01 | 0.001 | $10^{-4}$ | $10^{-5}$ |
|---|---|---|---|---|---|---|---|
| $\omega/k$ | 19948.238-394.85476 i | 18004.713-1134.8478 i | 17352.495-129.11251 i | -1.5286399e-07 + 14085.194 i | -2.7558417e-07 + 251882.17 i | -0.0014348603 + 2534299.1 i | 4.4096261e-05 + 25344553. i |
| $\tilde{\rho}_{1n}$ | 3.7185034e-05+ 7.3603934e-07i | 3.4553753e-05+ 2.1779437e-06i | 3.4274573e-05+ 2.5502246e-07i | -2.1650459e-16 - 4.5667966e-05i | 3.8460304e-17 - 2.8029347e-06i | 2.0321975e-16 - 2.7901044e-07i | 5.7319982e-20 - 2.7899749e-08i |
| $v_{1nx}$ | 0.74206653 | 0.62460204 | 0.59478227 | 0.64324215 | 0.70600927 | 0.70709591 | 0.70710667 |
| $v_{1ny}$ | 0.0059536450+ 0.059013026i | 0.27371929+ 0.19274930i | 0.36504460+ 0.023135908i | 0.28502434+ 2.0688807e-12i | 0.0086815943+ 6.5952894e-12i | 8.8051761e-05+ 1.1632423e-14i | 8.8064354e-07+ 9.9096360e-15i |
| $v_{1nz}$ | 7.2791467e-18+ 5.4067156e-17i | 1.7293074e-17+ 1.0037209e-16i | 8.5146807e-16 - 6.7902746e-15i | 3.3047214e-15+ 9.1759418e-18i | -1.3614275e-15+ 5.8706425e-15i | 6.9886484e-16 - 8.5668087e-17i | 5.6824800e-16 - 1.2735148e-26i |
| $\tilde{\rho}_{1i}$ | 2.9808444e-05+ 1.4807120e-06i | 3.2144409e-05+ 5.5994530e-06i | 3.4245653e-05+ 7.1132825e-07i | -2.4263513e-16 - 4.5632338e-05i | 2.9276760e-14 - 2.8028261e-06i | 1.3814852e-16 - 2.7901033e-07i | -2.7478735e-17 - 2.7899749e-08i |
| $v_{1ix}$ | 0.59521059 + 0.017767588i | 0.58510539 + 0.064337533i | 0.59433936 + 0.0079217776i | 0.64274033 + 4.3938811e-14i | 0.70598194 - 1.2106820e-14i | 0.70709563 - 1.4103302e-14i | 0.70710667 - 3.0511276e-14i |
| $v_{1iy}$ | 0.29966740 + 0.023496451i | 0.35271344 + 0.064074845i | 0.36593044 + 0.0072923535i | 0.28602800 + 2.0313569e-12i | 0.0087362628+ 6.6114148e-12i | 8.8609634e-05+ 4.1518037e-14i | 8.8622342e-07+ 9.9044818e-15i |
| $v_{1iz}$ | -2.7621934e-15+ | 6.0913648e-17 - 7.5869916e- | 4.8863072e-16 - 6.8986091e- | 3.2478650e-15+ 9.1704774e- | -1.3614216e-15+ | 6.9886351e-16 - 8.5667992e- | 5.6824799e-16 - 2.3840256e- |
|  | 6.4090085e-14i | 17i | 15i | 18i | 5.8706405e-15i | 17i | 24i |
| $b_{1y}$ | -0.0035535447 - 0.029371486i | -0.12867648 - 0.078994392i | -0.15843754 - 0.0088583721i | -1.9193072e-12 - 0.10036685i | 9.6285365e-12 - 0.054669166i | 2.8941444e-12 - 0.0055788080i | -5.4957458e-13 - 0.00055799497i |
| $b_{1z}$ | -1.1457310e-17 + 1.9002189e-17i | -1.3222028e-16 - 1.0615287e-16i | -8.4947394e-16 - 6.0189880e-17i | 2.3666421e-20 + 8.1368573e-17i | 1.6781742e-22 + 2.4921181e-18i | 7.9901794e-25 + 1.0769352e-17i | -3.0057546e-34 - 7.1647759e-26i |

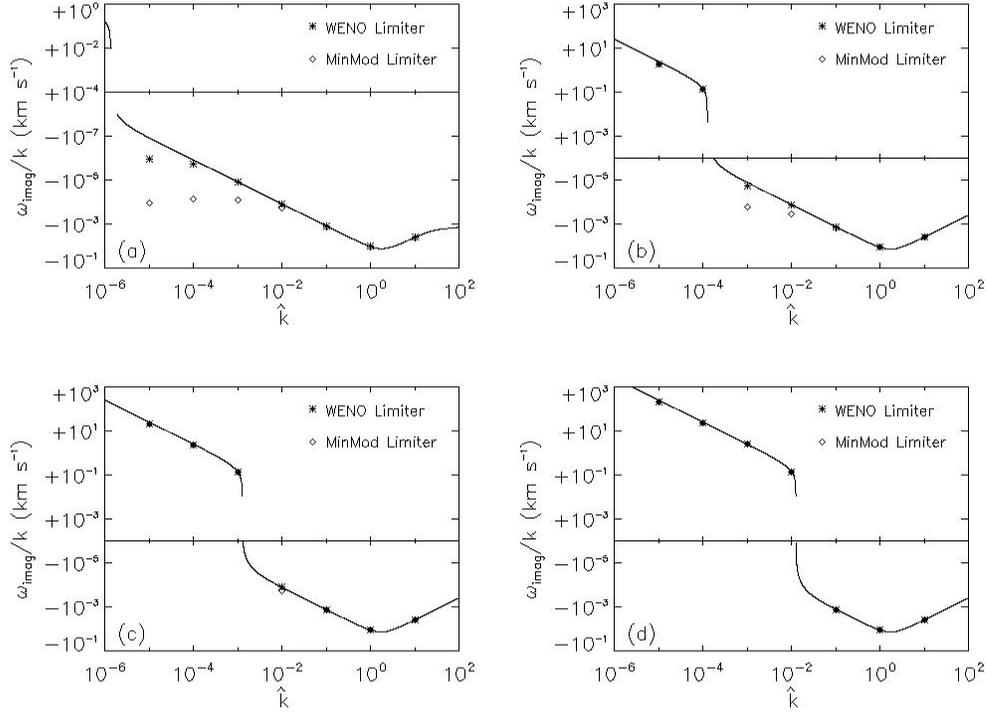

Figure 1: The magnitude of the imaginary component of the eigenvalues (lines) along with the measured magnitudes of the imaginary components from the *neutral* density in simulations at a resolution of 128 zones per wavelength, using a MinMod limiter (diamonds) and WENO limiter (stars). Fig. 1a displays the results for an ionization fraction of $10^{-2}$, Fig. 1b displays the results for an ionization fraction $10^{-4}$, Fig. 1c displays the results for an ionization fraction $10^{-5}$, and Fig. 1d displays the results for an ionization fraction of $10^{-6}$.

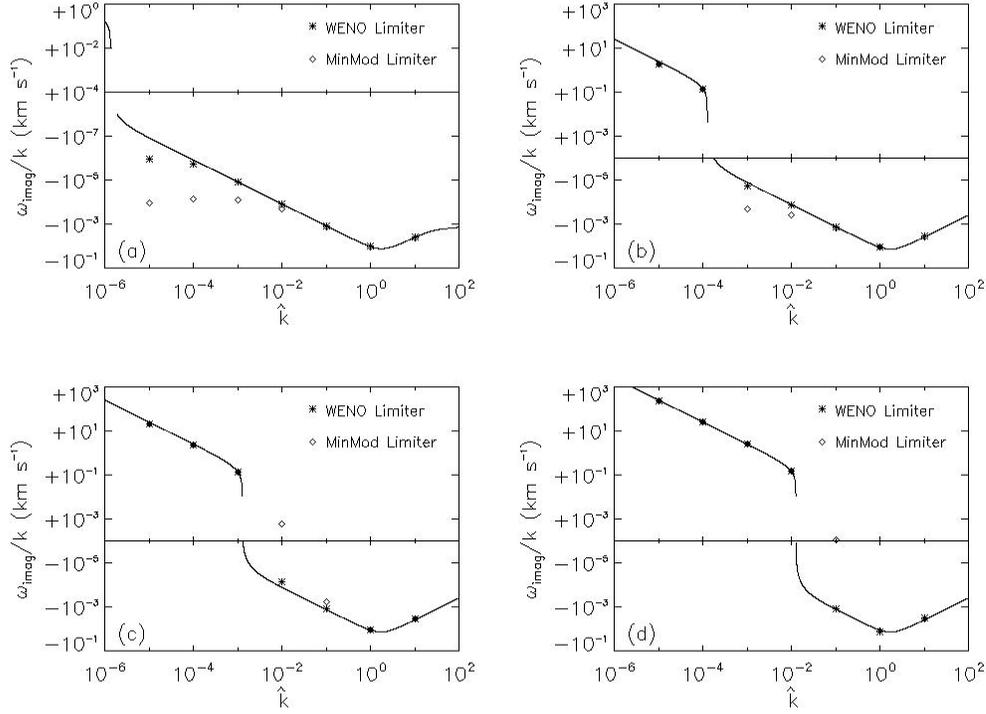

Figure 2: The magnitude of the imaginary component of the eigenvalues (lines) along with the measured magnitudes of the imaginary components from the *ion* density in simulations at a resolution of 128 zones per wavelength, using a MinMod limiter (diamonds) and WENO limiter (stars). Fig. 2a displays the results for an ionization fraction of $10^{-2}$, Fig. 2b displays the results for an ionization fraction $10^{-4}$, Fig. 2c displays the results for an ionization fraction $10^{-5}$, and Fig. 2d displays the results for an ionization fraction of $10^{-6}$.

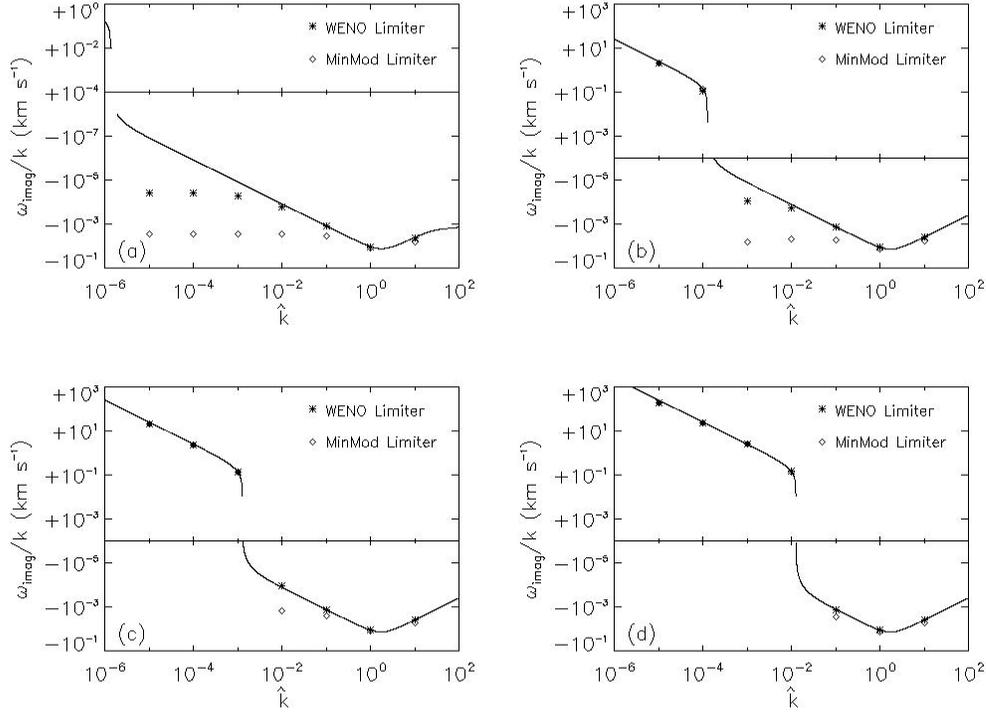

Figure 3: The magnitude of the imaginary component of the eigenvalues (lines) along with the measured magnitudes of the imaginary components from the *neutral* density in simulations with a resolution of 32 zones per wavelength, using a MinMod limiter (diamonds) and WENO limiter (stars). Fig. 3a displays the results for an ionization fraction of $10^{-2}$, Fig. 3b displays the results for an ionization fraction $10^{-4}$, Fig. 3c displays the results for an ionization fraction $10^{-5}$, and Fig. 3d displays the results for an ionization fraction of $10^{-6}$.

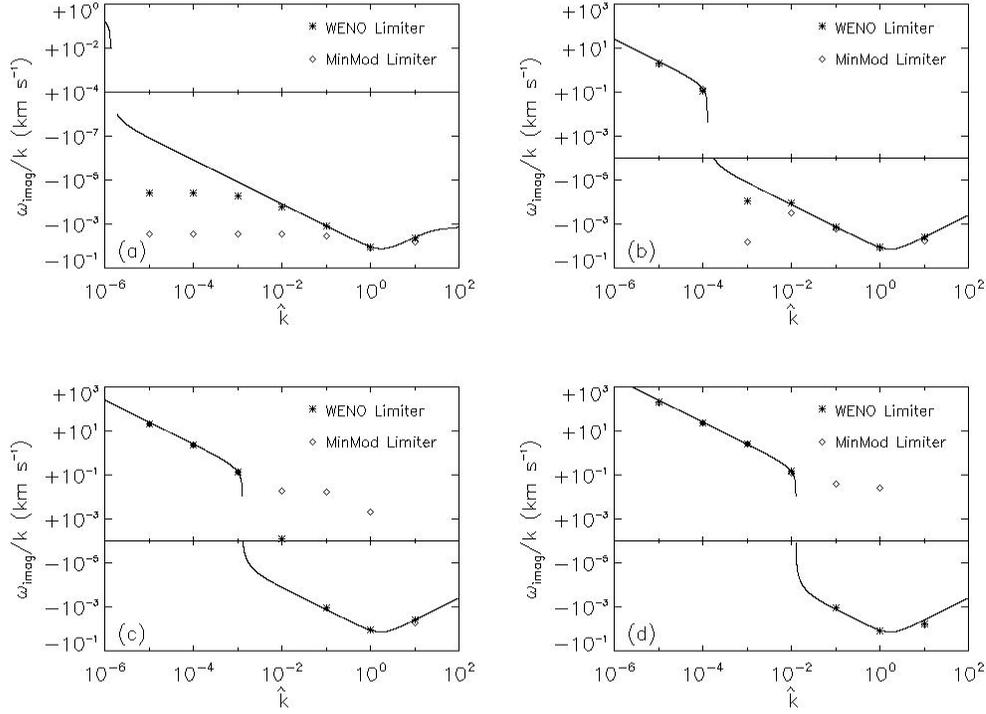

Figure 4: The magnitude of the imaginary component of the eigenvalues (lines) along with the measured magnitudes of the imaginary components from the *ion* density in simulations at a resolution of 32 zones per wavelength, using a MinMod limiter (diamonds) and WENO limiter (stars). Fig. 4a displays the results for an ionization fraction of $10^{-2}$, Fig. 4b displays the results for an ionization fraction $10^{-4}$, Fig. 4c displays the results for an ionization fraction $10^{-5}$, and Fig. 4d displays the results for an ionization fraction of $10^{-6}$.

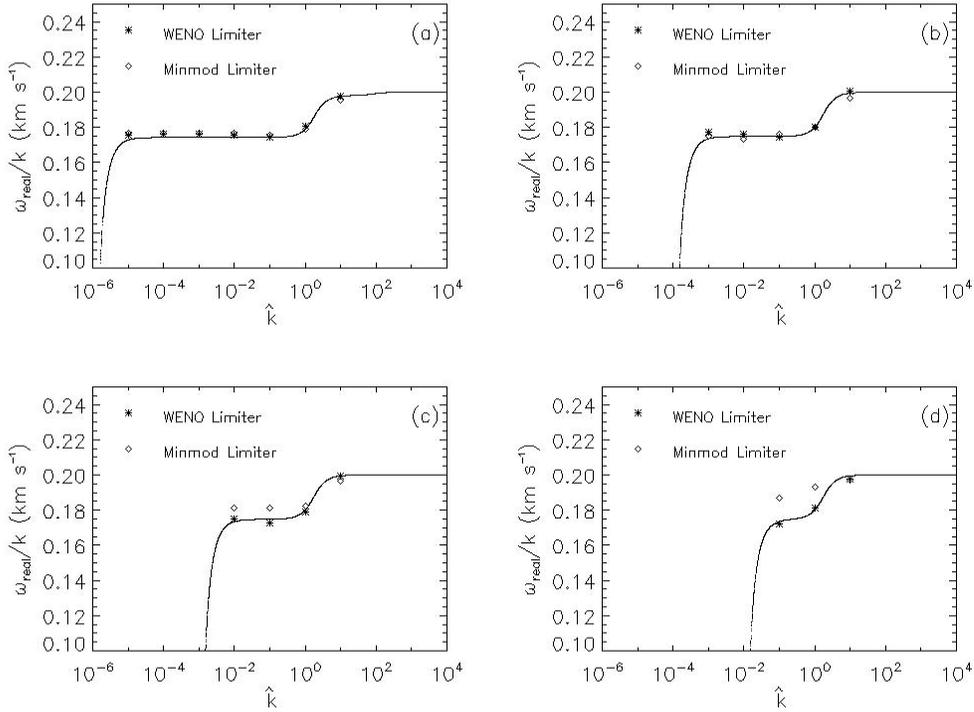

Figure 5: The phase velocity of the slow wave measured from our simulations at a resolution of 32 zones per wavelength, as measured by the x-velocity of the neutrals, using a MinMod limiter (diamonds) and a WENO limiter (stars). The solid line is the predicted real part of the eigenvalue. Fig. 5a displays the results for an ionization fraction of $10^{-2}$, Fig. 5b displays the results for an ionization fraction of $10^{-4}$, Fig. 5c displays the results for an ionization fraction of $10^{-5}$, and Fig. 5d displays the results for an ionization fraction of $10^{-6}$.

TWO-FLUID AMBIPOLAR DIFFUSION

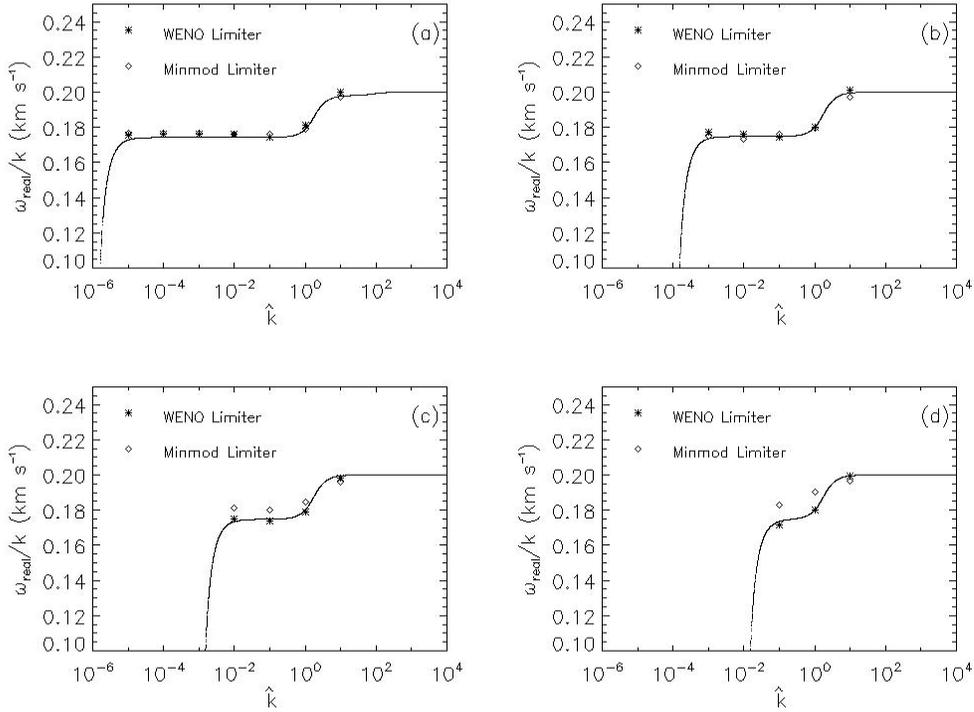

Figure 6: The phase velocity of the slow wave measured from our simulations at a resolution of 32 zones per wavelength, as measured by the x-velocity of the neutrals, using a MinMod limiter (diamonds) and a WENO limiter (stars). The solid line is the predicted real part of the eigenvalue. Fig. 6a displays the results for an ionization fraction of $10^{-2}$, Fig. 6b displays the results for an ionization fraction of $10^{-4}$, Fig. 6c displays the results for an ionization fraction of $10^{-5}$, and Fig. 6d displays the results for an ionization fraction of $10^{-6}$.

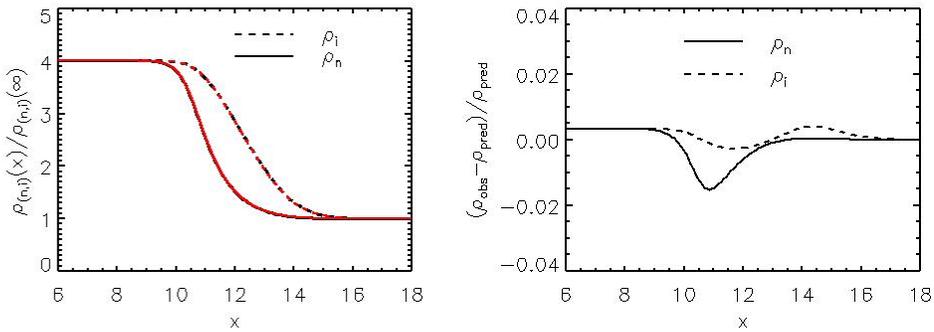

Figure 7: (a) The density structure of a C-shock (black lines), with the predicted solution (red lines), for both the neutral (solid line) and ionized (dashed line) fluids. For clarity, we show the portion of the grid that contains the shock; the boundaries at 0.0 and 20.0 mpc are not shown. (b) Error in the simulation for both the neutral and ionized fluids.

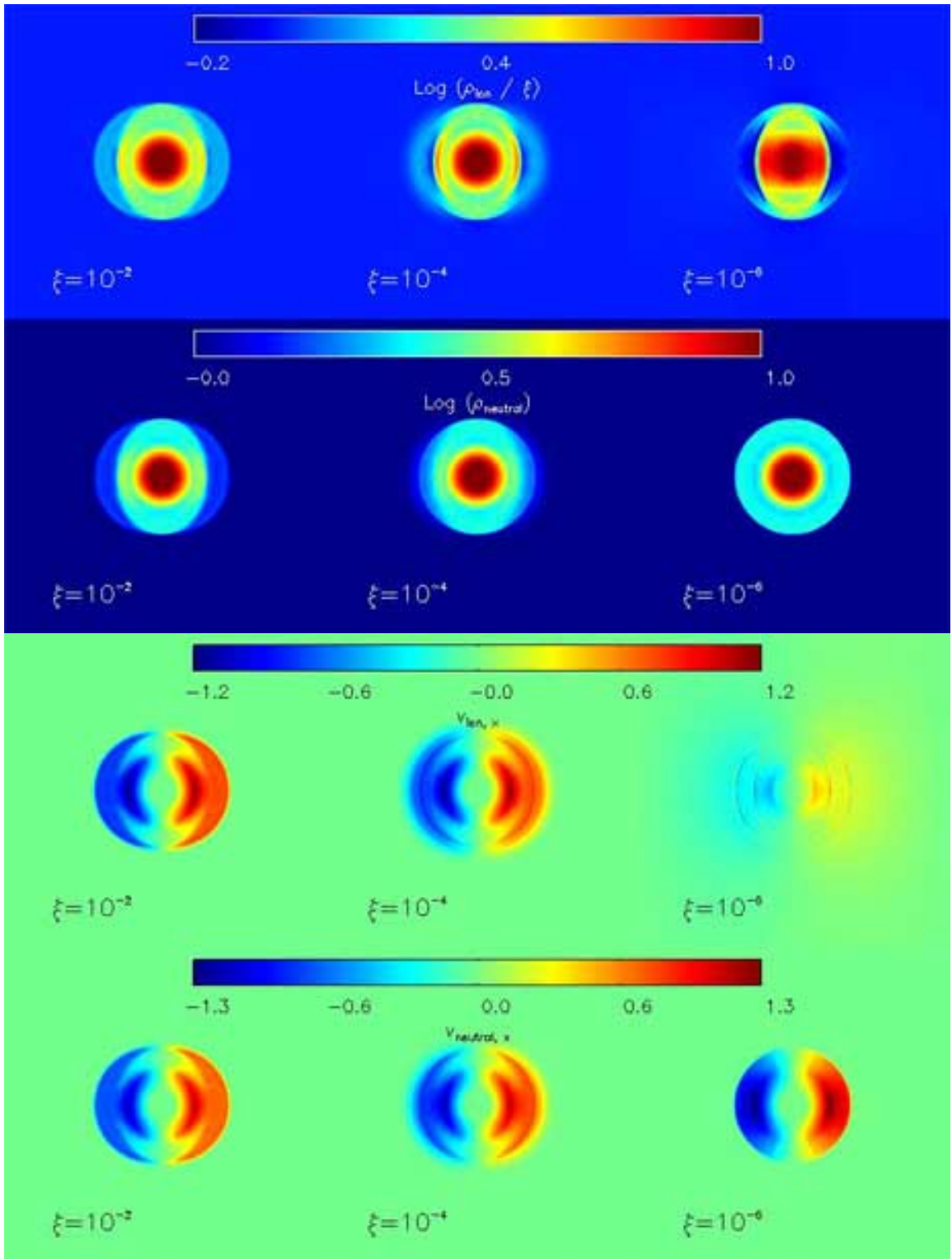

Figure 8. Density and velocity for the circular blast wave problem.

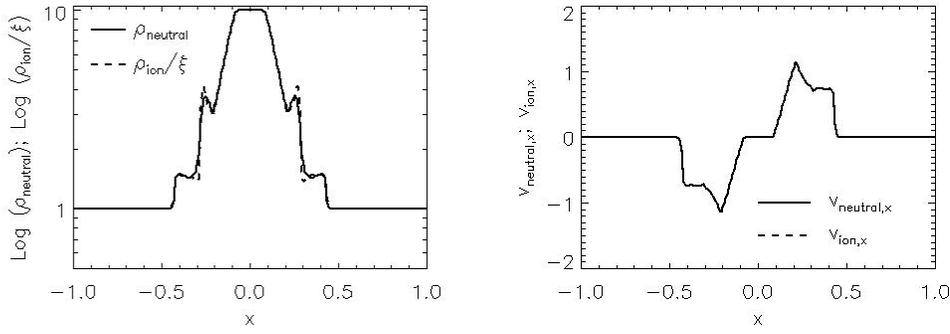

Figure 9. Cross-section through the midplane, perpendicular to the initial magnetic field, for the two-fluid shock pulse with an ionization fraction of $10^{-2}$. The densities of the ionized and neutral fluids are plotted on the left, and the x-velocities (perpendicular to the initial magnetic field) are plotted on the right. To show the neutral and ionized densities on the same scale, the ion densities have been divided by the ionization fraction.

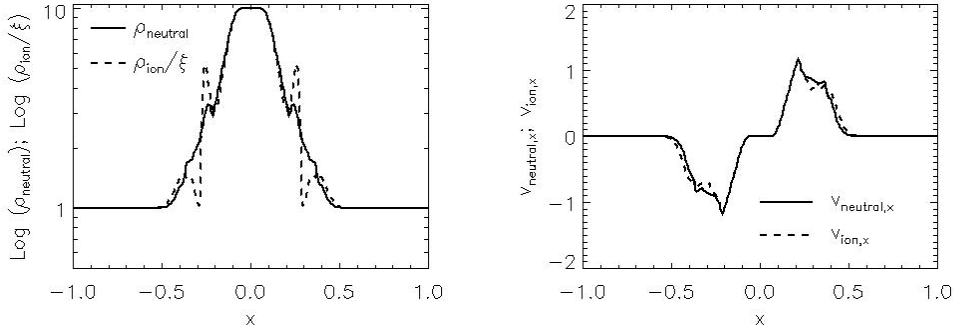

Figure 10. Cross-section through the midplane, perpendicular to the initial magnetic field, for the two-fluid shock pulse with an ionization fraction of $10^{-4}$. The densities of the ionized and neutral fluids are plotted on the left, and the x-velocities (perpendicular to the initial magnetic field) are plotted on the right. To show the neutral and ionized densities on the same scale, the ion densities have been divided by the ionization fraction.

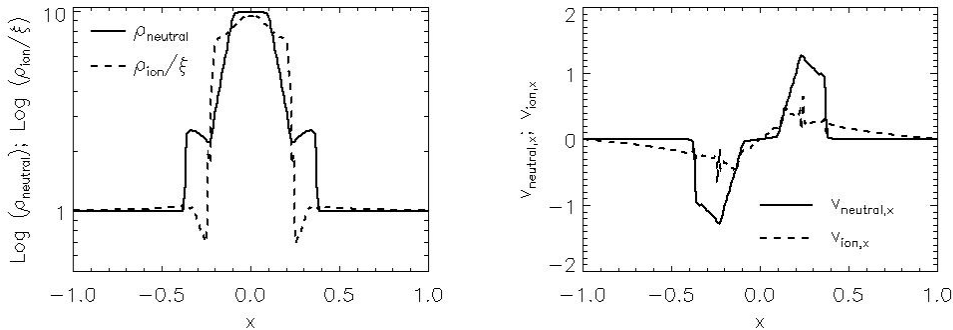

Figure 11. Cross-section through the midplane, perpendicular to the initial magnetic field, for the two-fluid shock pulse with an ionization fraction of $10^{-6}$. The densities of the ionized and neutral fluids are plotted on the left, and the x-velocities (perpendicular to the initial magnetic field) are plotted on the right. To show the neutral and ionized densities on the same scale, the ion densities have been divided by the ionization fraction.

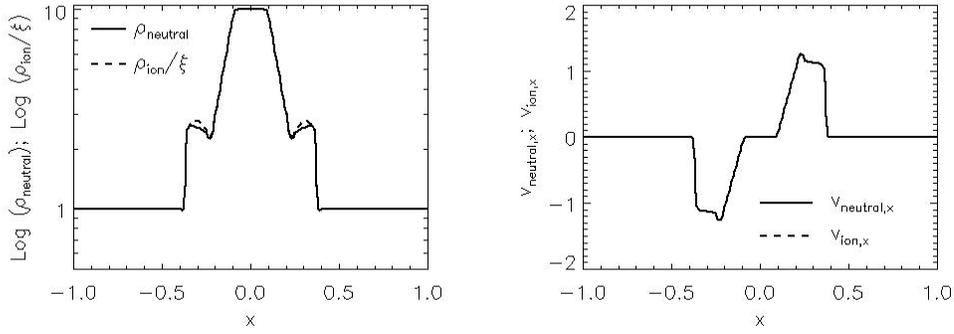

Figure 12. Cross-section perpendicular to the midplane, parallel to the initial magnetic field, for the two-fluid shock pulse with an ionization fraction of $10^{-2}$. The densities of the ionized and neutral fluids are plotted on the left, and the y-velocities (parallel to the initial magnetic field) are plotted on the right. To show the neutral and ionized densities on the same scale, the ion densities have been divided by the ionization fraction.

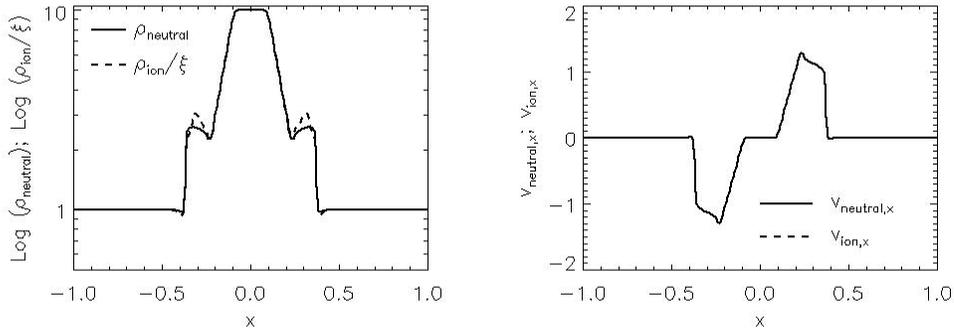

Figure 13. Cross-section perpendicular to the midplane, parallel to the initial magnetic field, for the two-fluid shock pulse with an ionization fraction of $10^{-4}$. The densities of the ionized and neutral fluids are plotted on the left, and the y-velocities (parallel to the initial magnetic field) are plotted on the right. To show the neutral and ionized densities on the same scale, the ion densities have been divided by the ionization fraction.

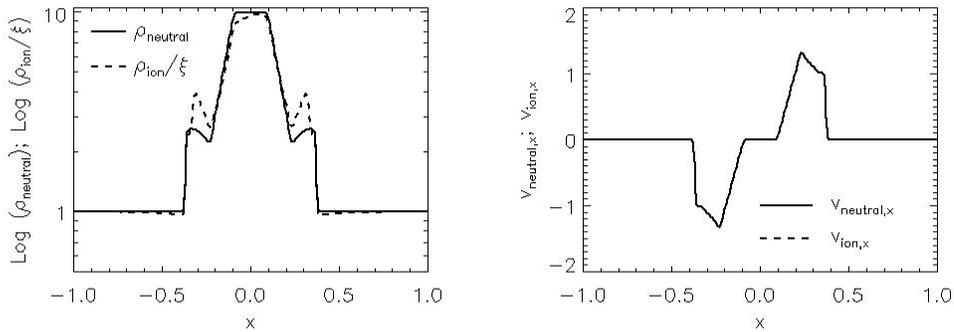

Figure 14. Cross-section perpendicular to the midplane, parallel to the initial magnetic field, for the two-fluid shock pulse with an ionization fraction of $10^{-6}$. The densities of the ionized and neutral fluids are plotted on the left, and the y-velocities (parallel to the initial magnetic field) are plotted on the right. To show the neutral and ionized densities on the same scale, the ion densities have been divided by the ionization fraction.

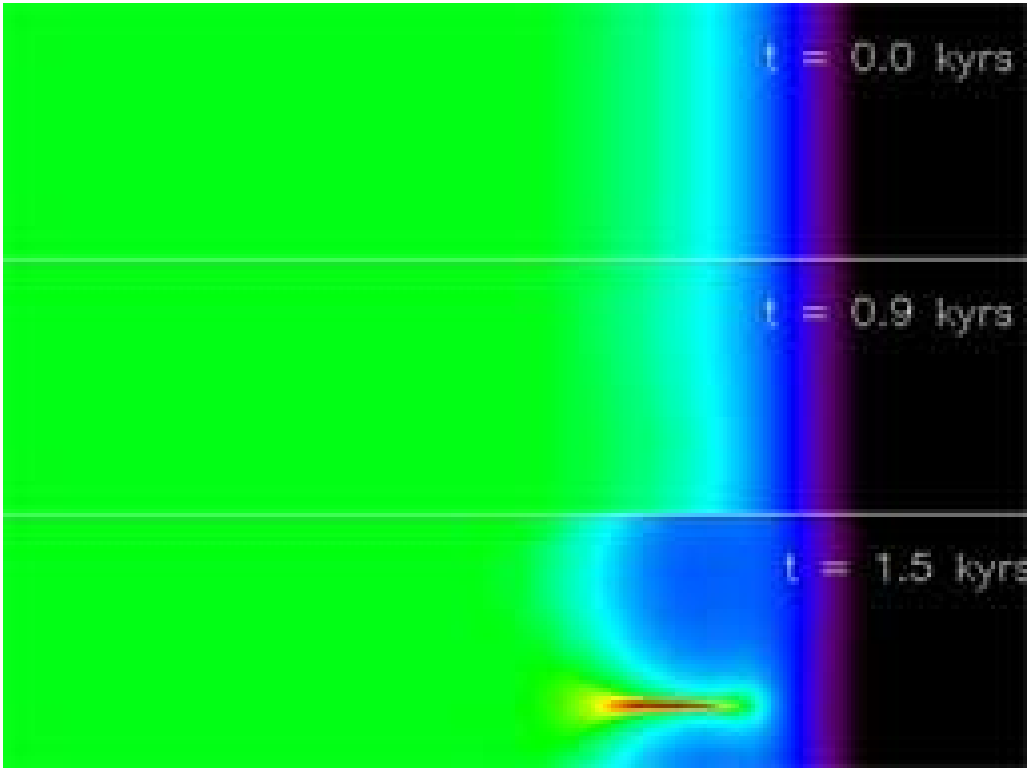

Figure 15: Logarithm of the ionized fluid density at 0.0, 0.9 and 1.5 kyrs.

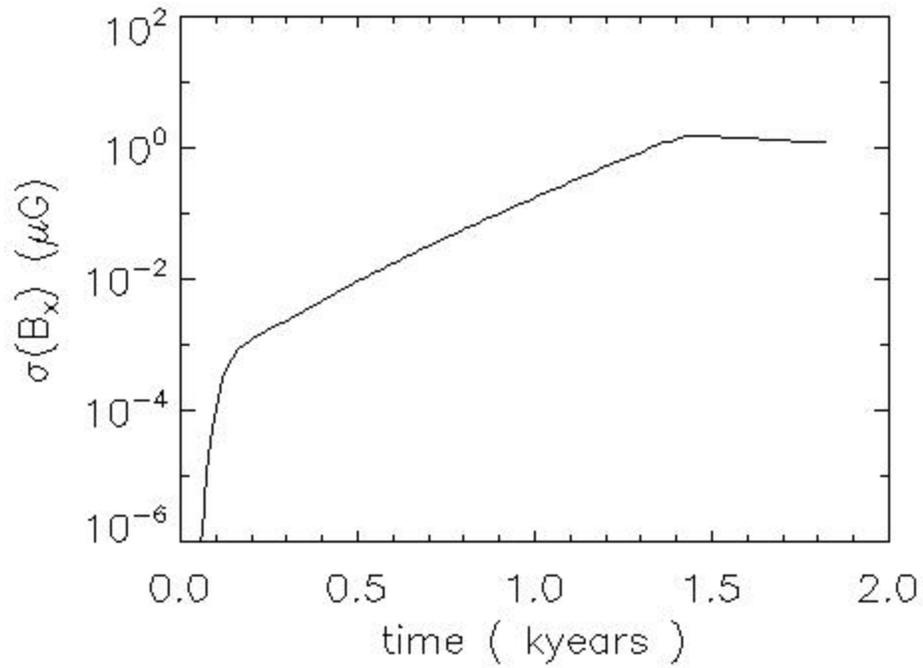

Figure 16: Growth of the instability, as measured by the standard deviation of the x-component of the magnetic field.